\documentclass[aps,pre,twocolumn,showpacs,floatfix]{revtex4}
\usepackage{epsfig}
\usepackage{psfrag}
\usepackage{color}
\definecolor{white}{rgb}{1,1,1}
\begin{document}
\title{Scaling behavior of fragment shapes}

\author{F.\ Kun${}^{1}\footnote{Electronic
address:feri@dtp.atomki.hu}$, F.\ K.\ Wittel${}^2$, H.\ J.\
Herrmann${}^3$, B.\ H.\ Kr\"oplin${}^2$, and K. J.\ M{\aa}l{\o}y${}^4$}
\affiliation{
${}^1$Department of Theoretical Physics,
University of Debrecen, P.\ O.\ Box:5, H-4010 Debrecen, Hungary \\
${}^2$ISD, University of Stuttgart, Pfaffenwaldring 27, D-70569
Stuttgart, Germany\\
${}^3$ICP, University of Stuttgart, Pfaffenwaldring 27, D-70569
Stuttgart, Germany\\
${}^4$Department of Physics, University of Oslo, P.O.Box:1048 -
Blindern, 0316 Oslo, Norway
}
\date{\today}

\begin{abstract}
We present an experimental and theoretical study of the shape of
fragments generated by explosive and impact loading of closed
shells. Based on high speed imaging, we have determined the fragmentation
mechanism of shells. Experiments have shown that the fragments vary from
completely isotropic to highly anisotropic elongated shapes, depending
on the microscopic cracking mechanism of the shell-material. Anisotropic
fragments proved to have self-affine character described by a scaling
exponent. The distribution of fragment shapes exhibits a
power law decay. The robustness of the scaling laws is illustrated by a
stochastic hierarchical model of fragmentation. 
Our results provide a possible improvement of the representation of
fragment shapes in models of space debris.
\end{abstract}


\maketitle
Spacecrafts and satellites, during their mission and
service time, are exposed to the danger of impact with pieces of space
debris, which is a growing population of rocket bodies,
non-functioning spacecrafts, rocket fuel ejecta and pieces of
fragmented material accumulated during 40 years of space exploration
\cite{debris_nasa_1997}.  
In order to minimize the potential hazard, objects of size larger than
10 cm are continuously tracked in space and their orbits are
taken into account for space activities. 
Fragmentation events like on-orbit explosions of fuel containers of
upper rocket stages and secondary breakups of fragments due to mutual
collisions, are the main source of the proliferation
of space debris, creating a large number of small
fragments which cannot be tracked. 
For safety reasons it is essential to work out models of
fragmentation, {\it i.e.} the breaking of objects into smaller
pieces, which are able to predict the consequences of on-orbit
explosions and impacts \cite{nasa_model_2001}. 
The NASA breakup model EVOLVE05 \cite{nasa_model_2001}, also implemented
by other space agencies, represents the fragments in terms
of their characteristic length $L_c$, surface-to-mass ratio $A/m$ and
velocity $\vec{v}$. Other quantities like the fragment
mass $m$ are determined from scaling relations. Model calculations are
performed in a 
phenomenological way, {\it i.e.} based on experiments and on-orbit
observations of breakup events, probability distributions of the above
quantities are prescribed. Monte Carlo simulations are carried
out taking into account the specific initial conditions of the event
studied \cite{nasa_model_2001}. 
The orbits of fragments are determined by their velocity, however, the 
lifetime of the orbits is limited by the atmospheric drag which mainly
depends on the shape of the fragments.  
The probability of impact of debris pieces with a spacecraft and the
resulting damage can be calculated from their velocity, mass, size, and
shape. The precision and 
predictive power of model calculations strongly rely on the quality of
the input distributions and the validity of scaling relations used.

General studies on fragmentation phenomena mostly
focused on the understanding of the mass distribution $F(m)$ of
fragments. For bulk solids, power law fragment mass
distributions $F(m) \sim m^{-\tau}$ have been obtained under widely
varying conditions with universal exponents $\tau$ depending mainly on the
spatial dimension $d$ 
\cite{turcotte_1986,glassplate_kadono_1997,kun_transition_1999,astrom_dynfrag_2004}.
Recently, we have pointed out that the fragmentation of shell-like objects,
like fuel containers or rocket bodies relevant for space debris, forms
an independent universality class \cite{shell_prl,shell_exp_imp},
which is also supported by other studies \cite{linna}.

\begin{figure}
\psfrag{aa}{\textcolor{white}{ a)}}
\psfrag{bb}{\textcolor{white}{ b)}}
\psfrag{cc}{\textcolor{white}{ c)}}
\psfrag{dd}{\textcolor{white}{ d)}}
\begin{center}
\epsfig{bbllx=0,bblly=0,bburx=610,bbury=160,file=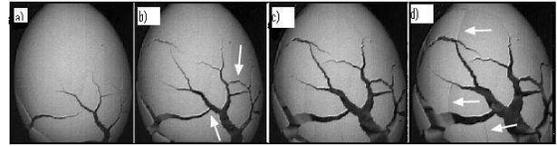,
  width=8.0cm}
 \caption{\small 
Consecutive snapshots of an exploding hen
egg-shell at a frame-rate $15000$ 1/sec.
Merging of side-branches of the hierarchical crack-tree and cracks
leading to secondary breakup are indicated by arrows in $b)$ and $d)$,
respectively.
}
\label{fig:egg_crack}
\end{center}
\vspace*{-0.8cm}
\end{figure}
In this paper we present a study of the
shape of fragments generated by explosive and impact loading of
closed shells. High speed imaging investigation of the explosion
process showed that the shape of fragments is determined by the
underlying cracking 
mechanism of shells which strongly depends on microscopic material
properties: fragments of isotropic shape are obtained for materials
where the branching-merging mechanism of cracks governs the breakup, 
while the formation of long straight cracks results in fragments of a
high degree of anisotropy. Fragments of anisotropic shape proved to
have self-affine character described by a scaling exponent. The
distribution of fragment shapes has a power law decay with a material
dependent exponent. We illustrate the robustness of the scaling laws
of the shape of shell fragments by a hierarchical stochastic model.
Our results suggest a possible improvement of the description of
fragment shapes in phenomenological breakup models of space debris by
clearly separating bulk and shell fragmentation and by taking into
account the effect of the cracking mechanism of different shell
materials on fragment shapes.

In order to understand how the breakup mechanism of shells determines
the shape of fragments, we have carried out explosion experiments of closed
shells made of brittle materials with different microscopic
properties such as hen and quail egg-shells and hollow glass spheres.
The egg-shells were used as a cheap and easy-to-handle
brittle bio-ceramics with a highly disordered microstructure
\cite{egg_icf11}. 
A Photron APX ultima high speed camera with  frame  rate 15000/s  and
spatial resolution $256\times 256$ pixels was used
to follow the time evolution of the explosion process, which also
enabled us to study the dynamics of crack formation on the surface,
and for the first time in the literature provided direct access to the
mechanism of fragmentation. 
The shells were filled with a stoichiometric hydrogen-oxygen mixture which was
electrically ignited approximately in the center of the shell. 
The analysis of the explosion of 20 egg-shells showed that the
breakup process starts with the 
nucleation of a few cracks at the flatter end of the egg (see Fig.\
\ref{fig:egg_crack}). Since the energy stored in the expanded shell at the 
instant of crack nucleation is high compared to the energy released by
the free fracture surface, the cracks propagate at a rather high
speed. The instability of the propagating crack results in sequential
splitting of the crack tip at almost regular distances, triggered by
the heterogeneities of the 
material (Fig.\ \ref{fig:egg_crack}$a$)
\cite{marder_physrep_1999}. 
\begin{figure}
\psfrag{aa}{a)}
\psfrag{bb}{b)}
\psfrag{cc}{c)}
\psfrag{dd}{d)}
\begin{center}
\epsfig{bbllx=0,bblly=0,bburx=600,bbury=145,file=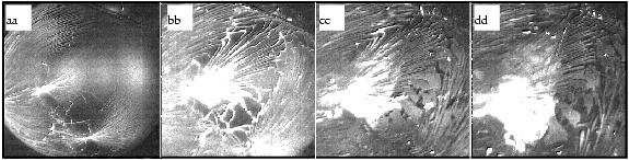,
  width=9.0cm}
 \caption{\small 
Consecutive snapshots of an exploding hollow glass sphere at a
frame-rate $15000$ 1/sec. {\it Hot spots} can be
seen in $a)$ radiating a large number of straight
cracks $b,c,d$.
}
\label{fig:glass_crack}
\end{center}
\vspace*{-0.8cm}
\end{figure}
The propagating sub-branches accelerate due to the expansion
of the shell and can undergo further splittings giving rise to a
hierarchical tree-like crack pattern (Fig.\ \ref{fig:egg_crack}$b$).
Fragments are formed along the main cracks by the merging of adjacent
side branches at almost right angles (Fig.\
\ref{fig:egg_crack}$c$). At most $3-4$
splittings can be observed along a main branch, however, merging
typically occurs at the first two levels of the hierarchy. 
We note that the area of
fragments generated at this stage does not show large variations, it
is practically determined by the inherent length scale of crack tip
splitting, furthermore, the shape of fragments is more or less isotropic.
The branching-merging process is initially governed by the in-plane
deformation 
of the shell, however, as the expansion increases
the out-of-plane deformation dominates
giving rise to further cracks mainly perpendicular to the former ones
(Fig.\ \ref{fig:egg_crack}$d$). This cracking
proceeds again in a sequential manner, typically breaking the fragments
into two pieces until a stable configuration is reached. 

The cracking mechanism discussed above should be generic to materials
with a strongly disordered microstructure
\cite{astrom_dynfrag_2004}. However, for shells made of 
materials like glass, the breaking mechanism can be significantly
different \cite{glassplate_kadono_1997,glass_crack_nature}.
Experiments on exploding hollow glass spheres have shown that the breakup
starts at one {\it hot spot} with random position on the surface (Fig.\
\ref{fig:glass_crack}$a$) from which long straight cracks 
radiate without any apparent branching (Fig.\
\ref{fig:glass_crack}$a,b$). 
This cracking mechanism results in a large number of long 
thin fragments having also a relatively large curvature which makes
them unstable against bending. During the expansion
of the sphere these primary fragments undergo a sequential breakup
process due to the out-of-plane bending deformation  (Fig.\
\ref{fig:glass_crack}$c,d$). Further details on the dynamics of
cracking of shells will be provided in Ref.\ \cite{wittel_unpub}.
Egg-shells and hollow glass spheres were also fragmented by impact
with a hard wall, which produced the same type of fragments
\cite{shell_exp_imp}. 

\begin{figure}
\begin{center}
\epsfig{bbllx=15,bblly=20,bburx=560,bbury=510,file=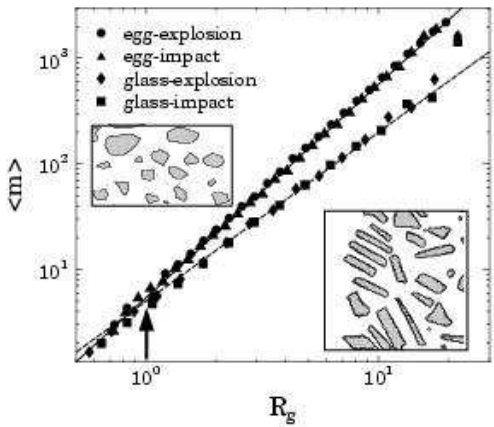,
  width=6.5cm}
 \caption{\small 
Log-log plot of $\left< m \right>$ as a function of $R_g$. The arrow
indicates the crossover point of glass fragments. Inset: scanned
pieces of egg (left) and glass (right).
}
\label{fig:mass_gyr}
\end{center}
\end{figure}

In the final state of the breakup process, the fragments
were carefully collected and digitized with a scanner for further
evaluation. It can be observed in Figs.\ \ref{fig:egg_crack},
\ref{fig:glass_crack} and in the inset of Fig.\ \ref{fig:mass_gyr}
that for the 
different types of materials considered,
the fragments are always compact two-dimensional objects with
little surface roughness, however, their overall shape can
vary from completely isotropic (egg-shell) to highly anisotropic (glass)
depending on the cracking mechanisms. 
The mass $m$ and surface $A$ of fragments is defined as the number of
pixels $N$ and the contour length of the spots in the digital image,
respectively.   
We characterize the linear extension of fragments by their
radius of gyration as $R_g^2 = (1/N)\sum_{i\neq
j=1}^N \left(\vec{r}_i-\vec{r}_j\right)^2$, where the sum goes
over the $N$ pixels $\vec{r}_i$ of the fragments.
In order to reveal how the shape of fragments varies with their
size, in Fig.\ \ref{fig:mass_gyr} the average fragment mass $\left< m
\right>$ is presented as a function of $R_g$ for different materials
from impact and explosion experiments. It is important to note that in
all cases power law functional forms $\left< m \right> \sim
R_g^\alpha$ are obtained with a high quality, 
however, the exponent $\alpha$ depends on
the structure of the crack pattern. Since 
the egg-shell pieces have regular isotropic shape,
their mass increases with the square of $R_g$ and hence $\alpha=2\pm
0.05$ was fitted. The large glass fragments are characterized by a 
significantly lower value of the exponent $\alpha = 1.5\pm 0.08$,
while for small glass pieces one observes a crossover to isotropic shape with
$\alpha = 2\pm 0.08$.
The value $\alpha < 2$ implies that the fragments have self-affine
character, {\it i.e.} the larger they
are, the more elongated they get. Note that similar anisotropy and
self-affinity of fragment shapes was not observed in $d$ dimensional
bulk fragmentation ($d=2, 3$)
\cite{astrom_dynfrag_2004,glassplate_kadono_1997,turcotte_1986,kun_transition_1999}.

To quantify this behavior, let us consider that the shell fragments
have a rectangular shape  
with side lengths $a$ and $b$, hence, the surface, mass, and radius of
gyration can be obtained as $A= 2(a+b)$, $m=ab$, and
$R_g=\sqrt{a^2+b^2}/(2\sqrt{3})$, respectively. 
The fragment mass can be expressed in terms of the aspect ratio
$r=a/b$ and $R_g$ as $m \sim R_g^2/\left(r+1/r\right)$, where even for
moderately elongated fragments the approximation $m \sim R_g^2/r$ is
valid. Consequently, for fragments with exponents $\alpha < 2$,
the aspect ratio must increase as a power of $R_g$ so that $r\sim
R_g^{\delta}$. Hence,
$m\sim R_g^{2-\delta}$ follows, and $\alpha = 2-\delta$, where $\delta \approx
1/2$ was obtained in the experiments. It is interesting to note
that the value $\delta = 1/2$ has been  
found in a broad class of systems producing self-affine structures,
for instance, for the scaling of the width with the length of the arms
of noise reduced DLA clusters, for clusters of directed percolation or for the
H\"older exponent of one-dimensional random walks \cite{dla}.

\begin{figure}
\begin{center}
\epsfig{bbllx=125,bblly=385,bburx=480,bbury=670,file=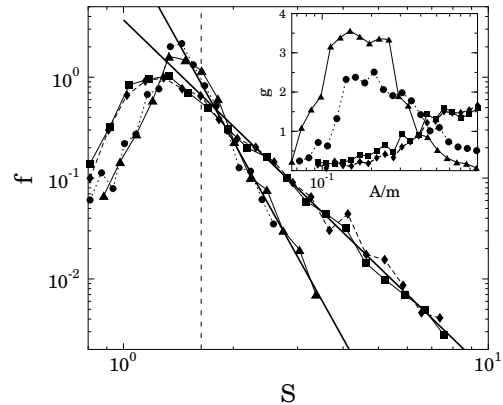,
  width=6.8cm}
 \caption{\small 
Distribution $f$ of the shape parameter $S$. Inset: distribution $g$
of $A/m$. The legend is the same as in Fig.\ \ref{fig:mass_gyr}.
}
\label{fig:shape_dist}
\end{center}
\end{figure}

Apart from the shape of individual fragments, it is also important to
know the probability of occurrence of a specific fragment shape in the
final state of a breakup process. The NASA breakup model characterizes the
shape of fragments by the surface-to-mass ratio $A/m$, the
distribution of which is fitted by a linear combination of Gaussian
distributions \cite{nasa_model_2001}. The functional form of the
corresponding distributions 
$g(A/m)$ of our shell pieces in the inset of Fig.\
\ref{fig:shape_dist} again shows a strong dependence on the cracking
mechanism. For isotropic fragments 
a reasonable fit could be obtained with Gaussians in agreement with
the NASA model \cite{nasa_model_2001}, however, for
anisotropic fragments $g$ increases monotonically.
The small sized
isotropic and the large very elongated anisotropic fragments both have large
$A/m$ value which prevents clear shape identification. 
To obtain a better characterization of 
fragment shapes, we introduce a dimensionless shape parameter
$S$ defined as $S = \frac{A}{m}R_g$,
multiplying the surface-to-mass ratio $A/m$ by the radius of gyration
$R_g$. Assuming rectangular objects, the shape parameter $S$ takes the
form $S = (a+b)\sqrt{a^2+b^2}/\left(\sqrt{3}ab\right)$. For fragments
of isotropic 
shape $a\approx b$, it follows that $S \approx 1.63$, which is
indicated by the vertical dashed line in Fig.\ \ref{fig:shape_dist}. If
the fragments are elongated $a \gg b$, the shape
parameter $S \approx a/b$ coincides with the aspect ratio $r$
characterizing the degree of anisotropy. 
Corresponding to the cracking mechanisms,
the distributions $f(S)$ of different materials and fragmentation
modes (explosion and impact) form two groups in Fig.\
\ref{fig:shape_dist}. Fragments of a low degree of anisotropy,
irrespective of their size, contribute to the maximum of $f(S)$ in
the vicinity of  $S\approx 1.63$.
Since egg-shell fragments are mostly isotropic
at all sizes, the distribution $f(S)$ decreases rapidly over
a narrow interval of $S$. The remarkable feature of the distribution
$f$ is that it follows a power law decay $f(S) 
\sim S^{-\beta}$, where the exponent $\beta=6.8\pm 0.3$ is obtained
for isotropic fragments over a limited scaling range, while $\beta
=3.5\pm 0.2$ follows when the cracking mechanism favors the formation
of anisotropic fragments. 

\begin{figure}
\begin{center}
\epsfig{bbllx=145,bblly=385,bburx=470,bbury=670,file=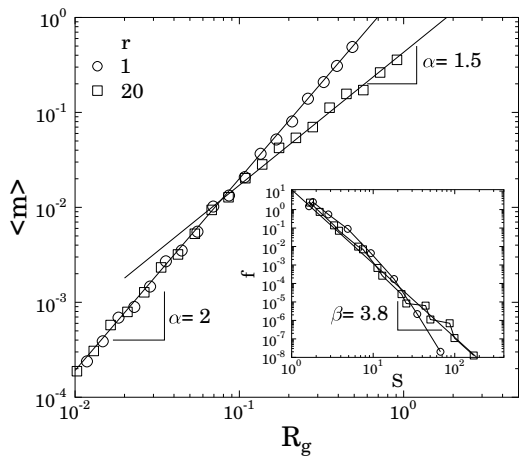,
  width=6.9cm}
 \caption{\small 
Log-log plot of $\left< m \right>$ as a function of $R_g$. Inset:
distribution $f$ of the shape parameter $S$.
}
\label{fig:model}
\end{center}
\end{figure}

Discrete element models of shell fragmentation usually consider highly
disordered brittle materials and provide isotropic fragment shapes,
but they have difficulties to capture microscopic mechanisms resulting
in long straight cracks \cite{shell_prl,shell_exp_imp}. 
We propose a simple stochastic binary breakup model in the spirit of
Refs.\  \cite{turcotte_1986,gonzalo_1995} to better understand the
experimental findings. 
The model focuses on the binary breakup of fragments formed by the
primary cracking mechanism of the shell.
Representing the fragments by rectangles with a continuous mass
distribution at a fixed aspect ratio $r=a/b$,
the effect of the material dependent primary cracking mechanism on the
shape of fragments can be taken into account by setting $r\approx 1$
for fragments of isotropic shape (egg), and $r \gg 1$ for the highly
anisotropic needle-like pieces (glass). Based on Figs.\
\ref{fig:egg_crack} and \ref{fig:glass_crack}, these fragments are
then assumed to undergo a sequential binary breakup 
process, where at each step of the hierarchy they break into
two pieces of equal mass with a probability $p \leq 1$. Note that the
fragments have $1-p$ chance to keep their actual size.
To capture the effect of out-of-plain deformations, we choose
a side of a rectangle to break with a probability proportional to its
length.
Computer simulations of the model were performed starting from a
continuous distribution of fragment sizes varying the initial
aspect ratio
$r$ to model different materials, while $p$ was fixed. The
hierarchical process was 
followed up to $n=30$ generations resulting in $\sim 10^7$ fragments
in the final state, where the mass $m$, the radius of gyration $R_g$
and the shape parameter $S$ of fragments were determined. It can be
observed in Fig.\ \ref{fig:model} that similar to the experiments,
for all values of $r$ the average fragment 
mass exhibits a power law dependence on the radius of
gyration $\left< m\right> \sim R_g^{\alpha}$. 
Starting the process
with isotropic shapes $r\approx 1$, the fragments remain
isotropic at all levels of the hierarchy implying $\alpha = 2\pm 0.05$
as it was observed for the egg pieces. Modelling glass
fragments by a high initial anisotropy $r \gg 1$, the crack mostly
occurs along the same side of the rectangle lowering $r$, however,
when the two sides 
become comparable $r$ fluctuates about one. Consequently, large
fragments are characterized by an exponent $\alpha$ significantly
lower than 2, while for small pieces a crossover occurs to
isotropic shape with $\alpha=2$. In the experiments we estimated the
initial aspect ratio of glass fragments from Fig.\
\ref{fig:glass_crack} to fall in the range of $15 \le r \le
35$. Simulations with these aspect ratios $r$ proved 
to provide values of the exponent $\alpha$ in the vicinity of
$1.5$. In Fig.\ \ref{fig:model} the results are presented for $r=20$
where $\alpha = 1.5\pm 0.06$ was obtained .
When the fragments initially have an isotropic shape, the hierarchical
process gives rise to a rapidly decreasing distribution of the shape
parameter $f(S)$ over a narrow range as it was observed 
for egg pieces. Starting the simulation with elongated
fragments ($r=20$ in Fig.\ \ref{fig:model}) the
distribution $f$ shows a power law decay  $f(S) \sim S^{-\beta}$ 
with an exponent $\beta = 3.8 \pm 0.3$ very close to the experimental
value of glass fragments. 
It is very important to notice that these results for the exponent
$\alpha$ and for the distribution of fragment shapes are practically
independent on $p$.

Summarizing, based on high speed imaging techniques we have determined
the fragmentation mechanism of closed shells:
after the material-dependent primary cracking mechanism
governed by the in-plane deformation of the shell, a hierarchical
secondary breakup process sets in due to out-of-plane
deformations. Contrary to bulk systems, the shape of shell fragments
shows large variations from completely isotropic to highly
anisotropic fragments depending on the primary cracking mechanism.
We pointed out that the anisotropic fragments have a
self-affine character with a scaling exponent
$\delta=1/2$. To give a quantitative characterization of fragment shapes
we proposed a shape parameter the distribution of which was found to
exhibit a power law decay. 
A hierarchical stochastic breakup model provided quantitative
agreement with the experimental findings,
which demonstrates the robustness of the scaling laws of fragment
shapes in shell fragmentation.
The results imply that the characterization of fragment shapes
in breakup models of space debris production should be improved by a clear
distinction of bulk and shell fragmentation and by using scaling laws
with exponents depending on the cracking mechanism of the
material.

The authors are grateful to H.\ Klinkrad and C.\ Wiedeman of ESA for valuable
discussions. This work was supported by the project SFB381. F.\ K.\
was supported by OTKA T049209, M041537 and by the Gy.\
B\'ek\'esi Foundation of HAS.

\end{document}